\documentclass{ws2}

\begin{document}

\title{$\pi\rho\omega$ vertex in nuclear matter}

\author{Anna Baran, Agnieszka Bieniek, and Wojciech Broniowski}

\address{The H. Niewodnicza{\'n}ski Institute of Nuclear Physics, 
ul. Radzikowskiego 152 PL-31342 Cracow, Poland}  

\maketitle

\abstracts{Medium modifications of
the $\pi \omega \rho$ vertex are analyzed in context of the
$\omega \to \pi^0  \gamma^\ast$ and $\rho \to \pi \gamma^\ast$
decays in nuclear matter. A relativistic hadronic model 
with mesons, nucleons, and $\Delta(1232)$ isobars is applied. 
A substantial increase of the widths for the decays
$\omega \to \pi^0 \gamma^\ast$ and $\rho \to \pi \gamma^\ast$ is found 
for photon virtualities in the range $0.3-0.6$GeV. This enhancement has
a direct importance for the description of dilepton yields obtained
in relativistic heavy-ion collisions.}

\noindent
{\em [To appear in Proceedings of the 2$^{\rm nd}$ International
Symposium on Quantum Theory and Symmetries,
Cracow, 18-21 July 2001, E. Kapu\'scik and A. Horzela, eds., 
World Scientific (Singapore, 2002)]}

\bigskip

The experience gathered over the past years led to the conviction that
the properties of hadrons in medium are substantially
modified \cite
{brscale,celenza,serot,chin,hatsuda,jean,herrmann1,pirner,rapp,%
LeupoldRev,Gao,hatlee,lee2000,Vogl,klingl,Chanfray,BH1,eletsky,%
FrimanActa,Lutz99}.
This view is supported by numerous studies of hadronic Green functions. 
Only a few
investigations of the in-medium hadronic {\em couplings} exist
\cite{Song2,BFH1,BFH2,Urban0,urban,Banerjee96,rhopipi}. 
For instance, it has been found that the value of the 
$\rho \pi \pi $ coupling is considerably enhanced.
In Ref. \cite{aa}, which is the basis for 
this talk, 
we have examined the {\em %
in-medium} $\pi \omega \rho $ coupling. This vertex enters the 
decays $\omega \rightarrow \pi ^{0}\gamma ^{\ast }$ and 
$\rho \rightarrow \pi \gamma^{\ast }$, which 
are important for description of the dilepton production 
in relativistic heavy-ion collisions \cite{ceres,helios}. The existing
calculations \cite{li,PKoch,rappadv} have always used the unmodified (vacuum) 
value of the $\pi \omega \rho $ coupling. We have found that the medium 
effects can significantly alter the input of such calculations, thereby
modifying the predicted dilepton production and creating hopes for a resolution
of the long-standing problem of the experimental dilepton enhancement in the 
range $M_{l^+l-}\sim 0.3-0.5$GeV.

In our study we use a typical hadronic model with meson, nucleon
and $\Delta$ degrees of freedom, and for simplicity 
work at leading baryon density and at zero
temperature. In addition, we constrain 
ourselves to the case where the decaying
particle is at rest with respect to the medium. The effects of the 
relative motion turn out to be not very large \cite{bbtbp}.
In the vacuum the $\pi \omega \rho$ coupling is 
\begin{equation}
-iV_{\pi \omega^\mu \rho^\nu }=i\frac{g_{\pi \omega \rho }}{F_{\pi }}%
\epsilon ^{\mu \nu p Q},  \label{vac}
\end{equation}
where we have used the notation $\epsilon ^{\mu \nu
pQ}=\epsilon ^{\mu \nu \alpha \beta }p_{\alpha }Q_{\beta }$, and
$F_\pi=93$MeV is the pion decay constant. In our convention 
$Q$ is the incoming momentum of the pion, $p$ is the outgoing momentum of
the $\rho$, and $q = Q-p$ is the outgoing momentum of the $\omega $. The vector
mesons are, in general, virtual. For
the kinematics where all external momenta
vanish, the vector meson dominance models gives $g_{\pi \to \omega \rho
}=-\frac{g_{\rho }g_{\omega }}{e^{2}}g_{\pi \gamma \gamma }$, with $g_{\pi
\gamma \gamma }=\frac{e^{2}}{4\pi ^{2}F_\pi}$. 
The nucleon propagator in nuclear matter is decomposed into the {\em %
free} and {\em density} parts in the usual way \cite{chin}: 
\begin{eqnarray}
iS(k) &=&iS_{F}(k)+iS_{D}(k)=i(\gamma ^{\mu }k_{\mu }+m_{N})[\frac{1}{%
k^{2}-m_{N}^{2}+i\varepsilon }+  \nonumber \\
&&\frac{i\pi }{E_{k}}\delta (k_{0}-E_{k})\theta (k_{F}-|k|)].  \label{S}
\end{eqnarray}
Here $m_{N}$ denotes the nucleon mass, $E_{k}=\sqrt{m_{N}^{2}+k^{2}}$, and $%
k_{F}$ is the Fermi momentum. The Rarita-Schwinger propagator \cite
{rarita,Ben} for the spin 3/2  $\Delta(1232)$ particle has the form 
\begin{equation}
iS_{\Delta }^{\mu \nu }(k)=i\frac{\gamma ^{\mu }k_{\mu }+M_{\Delta }}{%
k^{2}-M_{\Delta }^{2}}(-g^{\mu \nu }+\frac{1}{3}\gamma ^{\mu }\gamma ^{\nu }+%
\frac{2k^{\mu }k^{\nu }}{3M_{\Delta }^{2}}+\frac{\gamma ^{\mu }k^{\nu
}-\gamma ^{\nu }k^{\mu }}{3M_{\Delta }}).  \label{rarita}
\end{equation}
The effects of the non-zero width of the $%
\Delta $ are incorporated by replacing 
$M_{\Delta }$ by the complex mass $M_{\Delta }-i\Gamma /2$.

\begin{figure}[t]
\centerline{\psfig
{figure=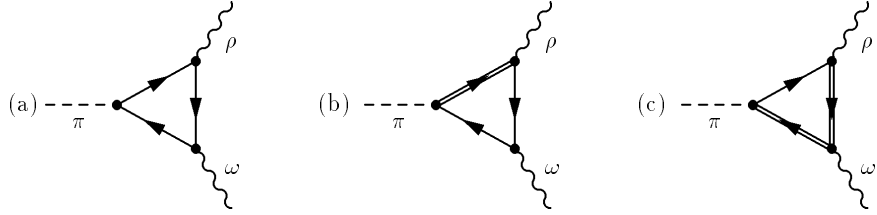,height=4.4cm,bbllx=75bp,bblly=377bp,bburx=541bp,bbury=537bp,clip=}}
\label{diag}
\caption{Diagrams included in our calculation (crossed ones not shown).
Wavy lines denotes the $\rho$ or $\omega$, 
dashed lines the pions, solid lines
the in-medium nucleon, and double lines the $\Delta$ resonance.}
\end{figure}

The diagrams used in our study are shown in Fig. 1.
The density effects are produced when exactly one of the nucleon lines 
in each of the our diagrams of Fig. 1 involves the nucleon 
{\em density} propagator, $S_{D}$. The diagrams with more than one 
$S_{D}$ vanish for kinematic reasons. On the other hand, diagrams with
no $S_{D}$  describe the vacuum polarization effects, not considered 
in the present study devoted to the Fermi sea effect only.

The meson-baryon vertices needed in our analysis have the 
standard form
\begin{eqnarray}
-iV_{\omega ^{\mu }NN} &=&ig_{\omega }\gamma ^{\mu }, \nonumber \\
-iV_{\rho _{b}^{\mu }NN} &=&i\frac{g_{\rho }}{2}(\gamma ^{\mu }-\frac{%
i\kappa _{\rho }}{2m_{N}}\sigma ^{\mu \nu }p_{\nu })\tau ^{b}, \\
-iV_{\pi ^{a}NN} &=&\frac{g_{A}}{2F_{\pi }}{Q}\!\!\!\!/ \gamma ^{5}\tau ^{a}.
\nonumber \end{eqnarray}
For the interactions involving the $\Delta$ resonance
we use
\begin{eqnarray}
-iV_{N\Delta ^{\alpha }\pi ^{a}} &=&g_{N\Delta \pi }Q^{\alpha }T^{a},
\nonumber  \\
-iV_{N\Delta ^{\alpha }\rho _{b}^{\mu }} &=&ig_{N\Delta \rho }({p}\!\!\!/%
\gamma ^{5}g^{\alpha \mu }-\gamma ^{\mu }\gamma ^{5}p^{\alpha })T^{b}, \\
-iV_{\Delta ^{\alpha }\Delta ^{\beta }\omega ^{\mu }} &=&-ig_{\omega
}(\gamma ^{\alpha }\gamma ^{\mu }\gamma ^{\beta }-\gamma ^{\beta }g^{\alpha
\mu }-\gamma ^{\alpha }g^{\beta \mu }+\gamma ^{\mu }g^{\alpha \beta }).
\nonumber
\end{eqnarray}
Above $T^{a}$ is the standard isospin $\frac{1}{2}\rightarrow
\frac{3}{2}$ transition matrix \cite{EW,Ben,Hemmert,Pascal,Haber}. 
We choose the following values for the  physical
parameters: $g_{\omega } =10.4$, $g_{\rho }=5.2$, $g_{A}=1.26$, 
$g_{N\Delta \pi } =2.12/m_{\pi }$,
$g_{N\Delta \rho }=2.12 \sqrt{2}/m_{\pi }$, $\kappa _{\rho }=3.7$.  

The $\pi \omega \rho $ vertex has the following
tensor structure in the nuclear medium moving with the four-velocity $u$:
\begin{eqnarray}
A^{\mu \nu } &=&A_{1}\varepsilon ^{\mu \nu pQ}+A_{2}\varepsilon ^{\mu \nu
uQ}+A_{3}\varepsilon ^{\mu \nu pu}+A_{4}\varepsilon ^{\mu upQ}p^{\nu }
\label{struct} \\
&&+A_{5}\varepsilon ^{\mu upQ}Q^\nu+A_{6}\varepsilon ^{\mu upQ}u^{\nu
}+A_{7}\varepsilon ^{\nu upQ}p^{\mu }+A_{8}\varepsilon ^{\nu upQ}Q^{\mu
}+A_{9}\varepsilon ^{\nu upQ}u^{\mu }.  \nonumber
\end{eqnarray}
This structure, restricted by the Lorentz invariance and parity, is
more involved
than in the vacuum case (\ref{vac}) due to the availability of 
the additional four-vector $u$. The leading-density calculation of diagrams
of Fig. 1 is very simple, and follows the steps of Ref. \cite{rhopipi}.
First, the trace factors are evaluated, then $k^{\mu}$ is substituted by
$m_{N}u^{\mu }$, with $u^{\mu }=(1,0,0,0)$ in the rest frame of
the medium, and, finally, the loop integration is replaced by 
$\int \frac{d^{4}k}{(2\pi )^{4}}\Theta (k_{f}-\left| \vec{k}\right| )\delta
(k_{0}-E_{k})\rightarrow \frac{1}{8\pi }\rho _{B}$, where $\rho _{B}$ 
denotes the baryon density \cite{rhopipi}.

For the pion is at rest with respect to the medium,
the four-vectors $Q$ and $u$ are parallel, and then 
$A^{\mu \nu }=(A_{1}+A_{3}/m_{\pi})\varepsilon ^{\mu \nu pQ}$. 
We introduce the notation $A^{\mu \nu }=i\frac{e^{2}}{g_{\rho }g_{\omega }}%
\left( \frac{g_{\pi \rho \omega }}{F_{\pi }}+\frac{\rho _{B}}{F_{\pi }}%
B\right) \epsilon ^{\nu \mu pQ}$, which puts the medium effects in $B$.
Our results below will be presented
relative to the vacuum value $g_{{\rm vac}}=g_{\pi \rho \omega}$, hence we 
define $g_{{\rm eff}} \equiv g_{\pi \rho \omega }+\rho _{0}B$.  
 
For the processes $\omega \rightarrow \pi ^{0}\gamma ^{\ast }$ and $\rho
^{a}\rightarrow \pi ^{a}\gamma ^{\ast }$, where $\omega $ and $\rho $ are on
mass shell, we use the following vacuum values of the
coupling constants: $g_{\omega \rightarrow \rho \pi }=-1.13$ and $g_{\rho
\rightarrow \omega \pi }=-0.76$, obtained from the experimental partial
decay width $\Gamma _{\omega \rightarrow \pi ^{0}\gamma }=717{\rm keV}$ and $%
\Gamma _{\rho^a \rightarrow \pi^a \gamma }=79{\rm keV}$. These
coupling constants differ from each other, as well as from $g_{\pi
\rightarrow \omega \rho }=-\frac{g_{\rho }g_{\omega }}{e^{2}}g_{\pi \gamma
\gamma }=-1.36$ inferred from the $\pi ^{0}$ decay.
This difference is a manifestation
of form-factor effects related to virtuality of vector mesons.

The numerical results are shown in Fig. 2.
At low values of $M$ the
effective coupling constant remains practically unchanged. 
At higher values of $%
M$, above 0.2GeV, the value of $g_{{\rm eff}}$ is much 
larger than in the vacuum: for the square of $g_{\rm eff}$ we have an
enhancement by a factor of $\sim 2$ for the $\omega$, and $\sim 5$ 
for the $\rho$ around $M=0.4$GeV. 
These values strongly
indicate the possibility of large medium 
effects in the Dalitz decays of vector mesons.
\begin{figure}[tb]
\centerline{
\vspace{0mm} ~\hspace{-3.5cm} 
\epsfxsize = 10cm \centerline{\epsfbox{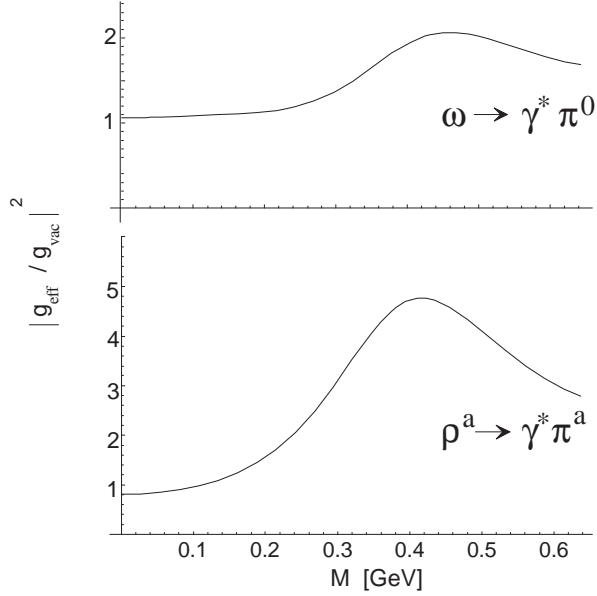}} \vspace{0mm}}
\label{fig:3}
\caption{Top: $|g_{{\rm eff}}/g_{{\rm vac}}|^{2}$ for the decay $%
\omega \to \gamma^\ast \pi^0$ at the nuclear saturation
density, plotted as a function of the virtuality of the photon, $M$. 
Bottom: the same for the decay $\rho$.}
\end{figure}
The enhancement of Fig. 2  directly affects the calculations of the
dilepton yields in relativistic heavy-ion collisions. 
The theoretical output
from the Dalitz decays of vector mesons becomes enhanced in the region
of $0.3-0.6$GeV, which is exactly where the existing calculations have
serious problems in supplying enough strength to explain the CERES and HELIOS
experiments \cite{ceres,helios}. The dilepton production rate from the Dalitz
decays of a vector meson is given by the formula \cite{PKoch} 
\begin{equation}
\frac{dN_{l^{+}l^{-}}}{d^{4}x\,dM^{2}}=\int \frac{d^{3}p_{v}}{(2\pi
)^{3}E_{v}}\,f(p_{v })\frac{3m_{v }}{\pi M^{3}}\Gamma
_{v \rightarrow \pi ^{0}\gamma ^{\ast }}\,\Gamma _{\gamma ^{\ast
}\rightarrow l^{+}l^{-}}  \label{PK}
\end{equation}
where $v$ relates to the vector meson, and $f(p_v)$ is the  
distribution function.
The width $\Gamma _{v \rightarrow \pi ^{0}\gamma ^{\ast }}$ depends on
the baryon density, $\rho _{B}$, which in turn, depends on the space-time
point $x$. Clearly,  an increased value of
$\Gamma _{\omega \rightarrow \pi ^{0}\gamma ^{\ast }}$
results in an increased dilepton yield.
A more exact estimate requires a detailed model of the evolution
of the fireball, and is left for a further study \cite{bbtbp}. 
According to our analysis ({\em cf.} Fig. 3), at $M\approx 400$MeV we have
 $\Gamma _{\omega \rightarrow \pi ^{0}\gamma ^{\ast }}\approx 900$keV
and $\Gamma _{\rho^a \rightarrow \pi^a \gamma ^{\ast }}\approx 250$keV. 
However,
in a thermalized system due to isospin degeneracy the $\rho $ mesons are
three times more numerous, hence the effect of the $\rho$ 
is comparable in size to the effect of the $\omega$.
The Dalitz yields from the $\pi ^{0}$ decays are not altered by the medium 
for a double reason: first,
virtually all pions decay outside of the
fireball due to their long lifetime, second, the width for this decay 
is practically unaltered by the medium \cite{aa}.

\section*{Acknowledgements}
Research supported by the Polish State Committee for
Scientific Research grant 2P03B09419.


\begin{thebibliography}{10}

\bibitem{brscale}
G.~E. Brown and M.~Rho, {\em Phys. Rev. Lett.}, 66:2720, 1991.

\bibitem{celenza}
L.~S. Celenza, A.~Pantziris, C.~M. Shakin, and W.-D. Sun,
{\em Phys. Rev.}, C45:2015, 1992.

\bibitem{serot}
B.~D. Serot and J.~D. Walecka, {\em Adv. Nucl. Phys.}, 16:1, 1986.

\bibitem{chin}
S.~A. Chin, {\em Ann. Phys. (NY)}, 108:301, 1977.

\bibitem{hatsuda}
T.~Hatsuda, H.~Shiomi, and H.~Kuwabara,
{\em Prog. Theor. Phys.}, 95:1009, 1996.

\bibitem{jean}
H.-C. Jean, J.~Piekarewicz, and A.~G. Williams,
{\em Phys. Rev.}, C49:1981, 1994.

\bibitem{herrmann1}
M.~Herrmann, B.~L. Friman, and W.~Noerenberg,
{\em Nucl. Phys.}, A560:411, 1993.

\bibitem{pirner}
B.~Friman and H.~J. Pirner,
{\em Nucl. Phys.}, A617:496, 1997.

\bibitem{rapp}
R.~Rapp, G.~Chanfray, and J.~Wambach,
{\em Nucl. Phys.}, A617:472, 1997.

\bibitem{LeupoldRev}
S.~Leupold and U.~Mosel,
{\em Prog. Part. Nucl. Phys.}, 42:221, 1999.

\bibitem{Gao}
{S. Gao, C. Gale, C. Ernst, H. Stocker, and W. Greiner},
{\em Nucl. Phys.}, A661:518, 1999.

\bibitem{hatlee}
T.~Hatsuda and S.~H. Lee, {\em Phys. Rev.}, C46:R34, 1993.

\bibitem{lee2000}
S.-H. Lee,
{\em Nucl. Phys.}, A670:119, 2000.

\bibitem{Vogl}
U.~Vogl and W.~Weise,
{\em Prog. Part. Nucl. Phys.}, 27:195, 1991.

\bibitem{klingl}
F.~Klingl, N.~Kaiser, and W.~Weise,
{\em Nucl. Phys.}, A624:527, 1997.

\bibitem{Chanfray}
G. Chanfray,
{\em Nucl. Phys.}, A685:328, 2001.

\bibitem{BH1}
W. Broniowski and B. Hiller,
{\em Phys. Lett.}, B392:267, 1997.

\bibitem{eletsky}
V.~L. Eletsky, B.~L. Ioffe, and J.~I. Kapusta,
 {\em Eur. J. Phys.}, A3:381, 1998.

\bibitem{FrimanActa}
B. Friman, {\em Acta Phys. Polon.}, B29:3195, 1998.

\bibitem{Lutz99}
M.~Lutz, B.~Friman, and G.~Wolf, {\em Nucl. Phys.}, A661:526, 1999.

\bibitem{Song2}
C. Song and V. Koch, {\em Phys. Rev.}, C54:3218, 1996.

\bibitem{BFH1}
W. Broniowski, W. Florkowski, and B. Hiller,
{\em Acta Phys. Polon.}, B30:1079, 1999.

\bibitem{BFH2}
W. Broniowski, W. Florkowski, and B. Hiller,
 {\em Eur. Phys. J.}, A7:287, 2000.

\bibitem{Urban0}
M.~Urban, M.~Buballa, R.~Rapp, and J.~Wambach,
{\em Nucl. Phys.}, A641:433, 1998.

\bibitem{urban}
M.~Urban, M.~Buballa, R.~Rapp, and J.~Wambach,
{\em Nucl. Phys.}, A673:357, 2000.

\bibitem{Banerjee96}
M.~K. Banerjee and J.~A. Tjon,
{\em Phys. Rev.}, C56:497, 1997.

\bibitem{rhopipi}
W.~Broniowski, W.~Florkowski, and B.~Hiller,
nucl-th/0103027.

\bibitem{aa}
A.~Bieniek, A.~Baran, and W.~Broniowski, nucl-th/0106053.

\bibitem{ceres}
{CERES Collab., G. Agakichiev {\it et al.}},
{\em Phys. Rev. Lett.}, 75:1272, 1995.

\bibitem{helios}
{HELIOS/3 Collab., M. Masera {\it et al.}},
{\em Nucl. Phys.}, A590:93c, 1995.

\bibitem{li}
G. Q. Li, C.~M. Ko, and G.~E. Brown,
{\em Nucl. Phys.}, A606:568, 1996.

\bibitem{PKoch}
P. Koch, {\em Z. Phys.}, C57:283, 1993.

\bibitem{rappadv}
R.~Rapp and J.~Wambach,
{\em Adv. Nucl. Phys.}, 25:1, 2000.

\bibitem{bbtbp}
A.~Bieniek and W.~Broniowski, to be published.

\bibitem{rarita}
W.~Rarita and J.~Schwinger,
{\em Phys. Rev.}, 60:61, 1941.

\bibitem{Ben}
M.~Benmerrouche, R.~M. Davidson, and N.~C. Mukhopadhyay,
{\em Phys. Rev.}, C39:2339, 1989.

\bibitem{EW}
T.~O.~E. Ericson and W.~Weise,
{\em Pions and nuclei},
Clarendon Press, Oxford, 1988.

\bibitem{Hemmert}
T.~R. Hemmert, B.~R. Holstein, and J. Kambor,
{\em J. Phys. G}, G24:1831, 1998.

\bibitem{Pascal}
V. Pascalutsa and R. Timmermans,
 {\em Phys. Rev.}, C60:042201, 1999.

\bibitem{Haber}
H. Haberzettl, nucl-th/9812043.

\end{thebibliography}
\end{document}